# Closed-form formulae of hyperbolic metamaterial made by stacked hole-array layers working at terahertz or microwave radiation

Piyawath Tapsanit[1,a)], Masatsugu Yamashita[2,b)], Chiko Otani[2,c)], Sriprajak Krongsuk[3,d)], Chesta Ruttanapuna[1,e)]

[1]Department of Physics, Faculty of Science, King Mongkut's Institute of Technology Ladkrabang, Chalongkrung Road, Ladkrabang, Bangkok, 10520, Thailand

[2]RIKEN Center for Advanced Photonics, Sendai, 980-0845, Japan

[3]Department of Physics, Faculty of Science, Khon Kaen University, Khon Kaen 40002, Thailand

[a)]p_epsilon@yahoo.com, [b)]m-yama@riken.jp, [c)]otani@riken.jp, [d)]srikro@kku.ac.th

**Abstract**

A metamaterial made by stacked hole-array layers known as a fishnet metamaterial behaves as a hyperbolic metamaterial at wavelength much longer than hole-array period. However, the analytical formulae of effective parameters of a fishnet metamaterial have not been reported hindering the design of deep-subwavelength imaging devices using this structure. We report the new closed-form formulae of effective parameters comprising anisotropic dispersion relation of a fishnet metamaterial working at terahertz or microwave frequency. These effective parameters of a fishnet metamaterial are consistent with those obtained by quasi-full solutions using known effective parameters of a hole-array layer working at frequency below its spoof plasma frequency with the superlattice period much smaller than the hole-array period. We also theoretically demonstrate the deep-subwavelength focusing at λ/83 using the composite structure of a slit-array layer and a fishnet metamaterial. It is found that the focused intensity inside a fishnet metamaterial is several times larger than that without the fishnet metamaterial, but the transmitted intensity is still restricted by large-wavevector difference in air and a fishnet metamaterial. Our effective parameters may aid the next-generation deep-subwavelength imaging devices working at terahertz or microwave radiation.

**Introduction**

The Abbe's diffraction limit restricts the spatial resolution of an optical imaging to about a half of light wavelength [1]. This optical barrier is caused by positive dielectric constant of an isotropic medium in which light propagates. The closed equi-frequency contour (EFC) of such medium prohibits the propagation of large-wavevector waves that encode subwavelength spatial resolution [2]. The diffraction limit can be beaten in an anisotropic medium having different signs of dielectric constants along an optical axis denoted as $\varepsilon_z$ and a transverse axis denoted as $\varepsilon_x$. This strong anisotropy results in hyperbolic EFC which lends such medium a name hyperbolic medium. The opened EFC of the hyperbolic medium allows the propagations of large-wavevector waves thereby surpassing the diffraction limit [3]. There are two types of hyperbolic medium: type-I with $\varepsilon_z < 0$, $\varepsilon_x > 0$, and type-II with $\varepsilon_z > 0$, $\varepsilon_x < 0$ [4]. Type-I hyperbolic metamaterial (HMM) has been applied to make a hyperlens [5], and type-II HMM finds the application in subwavelength focusing [6]. Both types have been naturally found in hexagonal boron nitride working in mid-infrared band [7]. However, the hyperbolic medium must be artificially fabricated in other frequency bands, the structure known as a hyperbolic metamaterial (HMM). In near-infrared to UV bands, the HMM is made by alternating metal and insulator layers [6,8]. The effective medium approximation (EMA) yields the effective parameters of this structure when metal and insulator thicknesses are much smaller than the working wavelength [9]. The effective parameters of the HMM reveal that the HMM's negative response comes from the negative dielectric constant of the metal below the metal's plasma frequency. However, in terahertz band whose non-ionizing energy is promising for non-invasive biomedical imaging [10,11], the metal behaves like perfect electrical conductor (PEC) and the alternating metal and insulator layers cannot be applied to these frequencies. J. B. Pendy et al. have shown that metallic hole-array exhibits spoof plasma frequency at the hole's cutoff frequency [12]. The stacked hole-array layers, known as fishnet metamaterial, has been employed to make low-loss negative-refractive-index metamaterial in near-infrared and visible bands working at wavelength slightly longer than the hole-array period [13,14]. The fishnet metamaterial (FM) working at wavelength much longer than the hole-array period behaves as hyperbolic medium as recently shown in microwave band [15], and near-infrared band [16]. However, to the best of our knowledge, the closed-form formulae of the effective parameters of the fishnet metamaterial, which are equivalent in principle to those derived from the EMA of alternating metal and insulator layers, have not been reported. The lack of these formulae hinders the design of deep-subwavelength imaging devices based on the FM that can work at terahertz or microwave frequency.

Here, we report the closed-form formulae of the effective parameters of the fishnet metamaterial working at terahertz or microwave band. The effective parameters are confirmed with quasi-full solutions (QFS) using homogenized metallic hole-array layer. We also theoretically demonstrate the subwavelength focusing using the hybrid structure of slit array and HMM.

**Method**

Effective parameters of a FM can be retrieved from zeroth-order reflection and transmission coefficients denoted as *r* and *t*, respectively, of a FM's unit cell, as shown in Fig. 1, by applying the following equation [16,17]

$$\cos(k_z L) = \frac{t^2 - r^2 + 1}{2t}, \quad (1)$$

where $k_z$ is longitudinal component of wavevector and *L* is FM thickness. Given frequency and transverse component of wavevector, the FM's EFC can be drawn by computing $k_z$ from Eq. (1) numerically. In Fig. 1, the metal layer as indicated by grey color is approximated as a PEC valid for terahertz or microwave radiations. Rectangular holes with hole width $b \geq a$ are periodically perforated onto the PEC layer with period *p* along *x* and *y* axes. An insulator layer with thickness *d* as indicated by blue color is combined to the hole-array layer with thickness *h* to form a superlattice with period $p_z=h+d$ along *z* axis. We are interested in deep-subwavelength scale, so we impose two assumptions: (1) $p \ll \lambda$, and (2) $b, a \ll \lambda$. The latter assumption leaves $TE_{01}$ waveguide mode denoted by $\langle \mathbf{r} | TE_{01} \rangle$ as the dominant mode inside the hole. Both assumptions have led to the introduction of spoof surface plasmons [12]. The structure is excited by transverse magnetic (TM) incident light whose electric field and wavevector lie on the *xz*-plane but magnetic field polarized along the *y* axis.

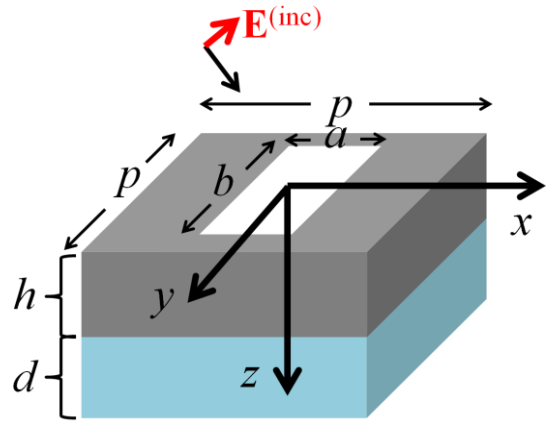

Fig. 1. Schematic view of a fishnet metamaterial's unit cell comprising metallic hole-array and insulator layers.

The transverse components of the incident electric and magnetic fields are given as follows

$$E_x^{(inc)} = E_{x0} e^{ik_z^{(in)}z} \langle \mathbf{r} | k_x, k_y \rangle, \quad H_y^{(inc)} = \frac{\omega \varepsilon_0}{k_0} Y_{00}^{(in)} E_{x0} e^{ik_z^{(in)}z} \langle \mathbf{r} | k_x, k_y \rangle, \quad (2)$$

where "inc" stands for incidence, "in" stands for input, $E_{x0}$ is amplitude of the incident electric field along *x* axis, $\omega$ is angular frequency, $k_0 = \omega/c$, $k_z^{(in)} = \sqrt{k_{in}^2 - k_x^2}$, $k_{in} = \sqrt{\varepsilon_{in}} k_0$ where $\varepsilon_{in}$ is dielectric constant of the input medium, $\langle \mathbf{r} | k_x, k_y \rangle = e^{ik_x x}/p$ is zeroth-order Bloch basis function, and $Y_{00}^{(in)} = \varepsilon_{in} k_0 / k_z^{(in)}$ is zeroth-order admittance of TM wave in the input medium. The impedance defined as $E_x / H_y$ is thus equal to $k_0 / \omega \varepsilon_0 Y_{00}^{(in)}$. The admittance in another region is obtained by changing dielectric constant and $z$-component of wavevector corresponding to that region. It will be later seen that a term containing the admittance corresponds to the coupling term. It should be noted that the Gauss's law is applied to obtain $E_z^{(inc)}$ from $E_x^{(inc)}$, and $H_y^{(inc)}$ is obtained from the electric field by applying the Faraday's law. The zeroth-order reflection and transmission coefficients are defined as fractions of $E_{x0}$ being reflected from and transmitted through the structure, respectively. Then, the $x$-component of the reflected electric field denoted by $E_x^{(r)}$, transmitted electric field denoted by $E_x^{(t)}$, and internal electric field in the insulator layer denoted by $E_x^{(d)}$ are expressed as follows

$$E_x^{(r)} = r E_{x0} e^{-ik_z^{(in)} z} \langle \mathbf{r} | k_x, k_y \rangle, \quad E_x^{(t)} = t E_{x0} e^{ik_z^{(out)}(z-p_z)} \langle \mathbf{r} | k_x, k_y \rangle, \tag{3}$$

$$E_x^{(d)} = \left( A e^{ik_z^{(d)}(z-h)} + B e^{-ik_z^{(d)}(z-h)} \right) \langle \mathbf{r} | k_x, k_y \rangle, \tag{4}$$

where $k_z^{(out)} = \sqrt{k_{out}^2 - k_x^2}$, $k_{out} = \sqrt{\varepsilon_{out}} k_0$, $\varepsilon_{out}$ is dielectric constant of the output medium, $k_z^{(d)} = \sqrt{k_d^2 - k_x^2}$, $k_d = \sqrt{\varepsilon_d} k_0$, and $\varepsilon_d$ is dielectric constant of the insulator. The remaining components are obtained by the $x$-component of the electric field in the same way as the incident light.

In couple-mode analysis [18], one writes $r$ in terms of the hole's field coefficient $E_{in}$, and $t$ in terms of the hole's field coefficient $E_{out}$ defined as follows

$$E_{in} = E_x^{(h)}(z=0), \quad E_{out} = -E_x^{(h)}(z=h), \tag{5}$$

where $E_x^{(h)}$ is $x$-component of TE$_{01}$ waveguide mode ($E_y^{(h)} = 0$). Then, $E_{in}$ and $E_{out}$ are found by solving two linear equations constructed by applying the continuities of $E_x$ and $H_y$ at interfaces in front of the hole-array layer and behind the hole-array layer. After applying these boundary conditions at $z=0$, the zeroth-order reflection coefficient can be written as

$$r = -1 + E_{in} s, \tag{6}$$

where $s = 4b \, \text{sinc}(k_x b/2) / p\pi\sqrt{2}$. Note that in the local model as adopted by J. B. Pendry et al [12], $\text{sinc}(k_x b/2) \approx 1$ and thus $s \approx 4b / p\pi\sqrt{2}$. We will also follow this model hereafter. We also have the linear equation of $E_{in}$ and $E_{out}$ written as follows

$$\left(iY_{00}^{(in)} - \gamma_{01}\right)E_{in}\langle k_x, k_y | TE_{01}\rangle - G_{01}^V E_{out}\langle k_x, k_y | TE_{01}\rangle = 2iY_{00}^{(in)}\langle k_x, k_y | k_x, k_y\rangle, \quad (7)$$

where $\gamma_{01} = \left(q_{01}^{(h)}/k_0\right)\cot\phi_{01}^{(h)}$, $G_{01}^V = \left(q_{01}^{(h)}/k_0\right)\csc\phi_{01}^{(h)}$, and $\phi_{01}^{(h)} = q_{01}^{(h)}h$. Eq. (7) is valid for any non-zero $\langle k_x, k_y |$ which can be taken out from the equation. By multiply the obtained equation by $\langle TE_{01}|$ and applying the orthonormality $\langle TE_{01}|TE_{01}\rangle$ excepting on the term having the admittance on the left hand side of the equation which is considered as the term describing the coupling of two waveguide modes via the zeroth-order diffraction mode [18], the first linear equation of $E_{in}$ and $E_{out}$ for the region in front of the hole-array layer is written as follows

$$\left(G_{in} - \gamma_{01}\right)E_{in} - G_{01}^V E_{out} = I_0, \quad (8)$$

where the input coupling parameter $G_{in} = iY_{00}^{(in)}s^2$ and $I_0 = 2iY_{00}^{(in)}s$. After applying the continuities of $E_x$ and $H_y$ at $z=h$ and $z=p_z$ and use the same coupling idea as in the previous case, the zeroth-order transmission coefficient and the second linear equation of $E_{in}$ and $E_{out}$ are written as follows

$$t = -\frac{\tau_{d,out}e^{i\phi_d}}{1+\rho_{d,out}e^{2i\phi_d}}E_{out}s, \quad (9)$$

$$-G_{01}^V E_{in} + \left(G_{out} - \gamma_{01}\right)E_{out} = 0, \quad (10)$$

where $\phi_d = k_z^{(d)}d$, $\tau_{\alpha,\beta}$ and $\rho_{\alpha,\beta}$ are transmission and reflection coefficients from media $\alpha$ to $\beta$ defined as follows

$$\tau_{\alpha,\beta} = \frac{2Y_\alpha}{Y_\alpha + Y_\beta}, \quad \rho_{\alpha,\beta} = \frac{Y_\alpha - Y_\beta}{Y_\alpha + Y_\beta}, \quad (11)$$

and $G_{out}$ is output coupling parameter defined as follows

$$G_{out} = iY_d\left(\frac{1-\rho_{d,out}e^{2i\phi_d}}{1+\rho_{d,out}e^{2i\phi_d}}\right)s^2 = G_{in}\frac{\alpha_d}{\alpha_{in}}, \quad (12)$$

where $\alpha_d = Y_d/\left(iY_{out}\sin\phi_d - Y_d\cos\phi_d\right)$, and $\alpha_{in} = Y_{in}/\left(iY_d\sin\phi_d - Y_{out}\cos\phi_d\right)$. By solving Eq. (8) and (10), we obtain $E_{in} = \left(G_{out} - \gamma_{01}\right)I_0/\Omega$ and $E_{out} = G_{01}^V I_0/\Omega$, where $\Omega = \left(G_{in} - \gamma_{01}\right)\left(G_{out} - \gamma_{01}\right) - G_{01}^{V^2}$, which then complete the solutions of $r$ and $t$. It should be noticed from the solution of $t$ that the resonant positions are now determined by both zeros of $\Omega$ as in the case of a bare hole-array layer and zeros of the denominator $1+\rho_{d,out}e^{2i\phi_d}$.

To obtain the effective parameters, the input medium is defined to be the same as the output medium. Then, with $H = q_{01}^{(h)} / k_{01}$ Eq. (1) can be written as follows

$$\cos(k_z p_z) = \frac{2\alpha_d^2 \alpha_{in} G_{in} H - \left[\alpha_d^2 G_{in}^2 / \alpha_{in} - \alpha_{in} H^2\right] \sin 2\phi_{01}^{(h)} + 2H\alpha_d G_{in} \cos 2\phi_{01}^{(h)}}{2\alpha_d \left[\left(\alpha_d G_{in}^2 - \alpha_{in} H^2\right) \sin \phi_{01}^{(h)} - H G_{in} \left(\alpha_{in} + \alpha_d\right) \cos \phi_{01}^{(h)}\right]}. \tag{13}$$

To simplify Eq. (13), we assume that $\left|\phi_{01}^{(h)}\right| \ll 1/2$ and $\left|\phi_d\right| \ll 1$ attainable by choosing thin hole-array and insulator layers whose thicknesses are much smaller than the working wavelength. These two assumption allow us to approximate $\sin \phi_{01}^{(h)} \approx \phi_{01}^{(h)}$, $\sin 2\phi_{01}^{(h)} \approx 2\phi_{01}^{(h)}$, and $\cos \phi_d \approx \cos 2\phi_d \approx 1$. Since only one unit cell is sufficient to retrieve the effective parameters, we can also approximate that $\cos(k_z p_z) \approx 1 - (k_z p_z)^2 / 2$. By substituting $\alpha_d$ and $\alpha_{in}$ into Eq. (13) and applying these approximations, Eq. (13) will become more complicated. Fortunately, we find that this complicated equation can be simplified by retaining only the most dominant terms. The simplified equation of Eq. (13) is written as follows

$$(k_z p_z)^2 = k_z^{(d)^2} d^2 + \frac{q_{01}^{(h)^2} k_z^{(d)^2} h d}{\varepsilon_d s^2 k_0^2}. \tag{14}$$

By substituting $q_{01}^{(h)} = \sqrt{\varepsilon_h k_0^2 - (\pi/b)^2}$ and $k_z^{(d)} = \sqrt{\varepsilon_d k_0^2 - k_x^2}$ into Eq. (14), we finally derive the anisotropic dispersion relation of the FM with effective parameters listed as follows

$$\frac{k_z^2}{\varepsilon_x} + \frac{k_x^2}{\varepsilon_z} = \mu_z k_0^2, \tag{15}$$

$$\mu_z = 1, \quad \varepsilon_z = \varepsilon_d, \quad \varepsilon_x = \left(\varepsilon_d f_d^2 + \frac{f_d f_h \varepsilon_h}{s^2}\right)\left(1 - f_p^2 / f^2\right), \tag{16}$$

where $f_h = h / p_z$ is filling ratio of hole-array layer, $f_d = 1 - f_h$, and $f_p$ is spoof plasma frequency defined as follows

$$f_p = \frac{c}{2b} \sqrt{\frac{f_h}{\varepsilon_d f_d s^2 + f_h \varepsilon_h}}. \tag{17}$$

These effective parameters indicate that the FM behaves like non-magnetic insulator material along $z$ axis, and behaves like a metal along $x$ axis with effective dielectric constant $\varepsilon_x$ written in the same form as the Drude's model. The FM is type-II hyperbolic medium at frequencies below the spoof plasma frequency in which $\varepsilon_x < 0$, but it becomes elliptical medium at frequencies above the spoof plasma frequency in which $\varepsilon_x > 0$. The spoof plasma frequency of the FM is lower than that of the bare hole-array layer, which is equal to $f_p = c/2b\sqrt{\varepsilon_h}$ [12], and both coincide when $f_h = 1$. For thin hole-array layer with $f_h \ll 1$, the spoof plasma frequency is approximately equal to $f_p \approx c\sqrt{f_h}/2bs\sqrt{\varepsilon_d}$. This means that the spoof plasma frequency can be lowered by decreasing $f_h$, and increasing $b$ and $\varepsilon_d$.

Another way to retrieve the effective parameters of a FM is by first homogenizing the hole-array layer with effective z-component of wavevector equal to $q_{01}^{(h)}$ and effective admittance equal to $\pi^2 p^2 q_{01}^{(h)}/8b^2 k_0$ [12]. Then, the transfer matrix method is applied to compute $r$ and $t$ of the composite structure comprising the homogenized hole-array layer and the insulator layer. Finally, the $r$ and $t$ coefficients are substituted into Eq. (1) to numerically compute $k_z$ for given $k_x$ and frequency. We compute $\varepsilon_x$ from the anisotropic dispersion relation with $\varepsilon_z = \varepsilon_d$, $\mu_z = 1$, $k_x = 0$ for a frequency above the spoof plasma frequency, and $k_x = \pi/p$ for a frequency below the spoof plasma frequency. The spoof plasma frequency is obtained at a frequency position at which real part of $\varepsilon_x$ is zero. The choice of $k_x$ will be later explained. This method will be called qausi-full solution (QFS) hereafter. The full details of QFS method are given in the supplementary material. To the best of our knowledge, this is the first time that QFS is applied to a FM. The QFS cannot explain the negative-refractive-index of a FM at a wavelength close to the hole-array period since it neglects diffraction modes $|m| \geq 1$ which contribute to the gap-spoof surface plasmons [13].

**Results**

1. **Effective parameters**

We first discuss the spoof plasma frequency $f_p$. The effective medium approximation (EMA) provides $f_p$ dependent on $f_h$ but not directly dependent on $h$ resulting from its assumption that $h \ll \lambda$. However, the QFS does not impose this assumption so that $f_p$ from QFS is expected to be dependent on $h$. To clarify when these two methods are consistent and different, we compare $f_p$ as a function of $f_h$ calculated by the EMA with those computed by QFS with three values of $h$ in Fig. 2(a). The $f_p$ from EMA as labelled by the black dashed line is perfectly consistent with the $f_p$ from QFS with $h=0.01p$ as labelled by the blue solid line. This is also true for thinner $h$ (not shown). However, the $f_p$ from QFS with $h=0.05p$ and $h=0.1p$, as labelled by green solid line and red solid line, respectively, become lower than the $f_p$ from EMA. Figure 2(b) shows the percentage difference of the $f_p$ between both methods with respect to the $f_p$ from EMA. This figure clearly shows that the $f_p$ from both methods are consistent when $h$ is small and $f_h$ is large. However, the $f_p$ from EMA will be much larger than that from QFS when $h$ is large and $f_h$ is small. Therefore, we must choose the supperlattice period $p_z$ determined by the ratio $h/f_h$ to be equal to or smaller than $0.01p$ in order to correctly apply the EMA.

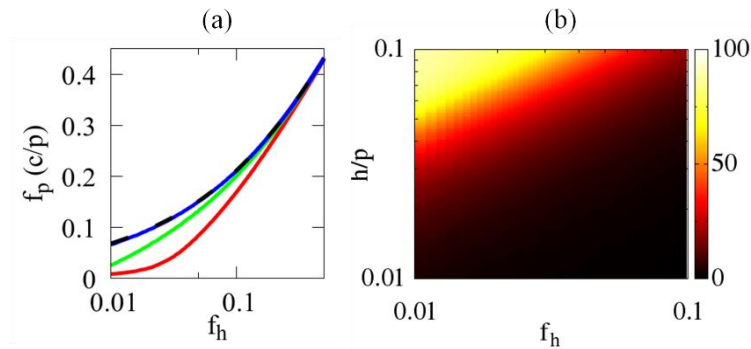

Fig.2. (a) Spoof plasma frequency as a function of $f_h$ computed by EMA (black dashed line) and by QFS with $h=0.01p$ (blue solid line), $h=0.05p$ (green solid line), and $h=0.1p$ (red solid line). (b) Percentage difference of the spoof plasma frequency from QFS with respect to that from EMA as a function of $f_h$ and $h$. Black and white colors represent 0% and 100%, respectively. The parameters of the FM's unit cell are $b=a=0.9p$, and $\varepsilon_h = \varepsilon_d = 1$.

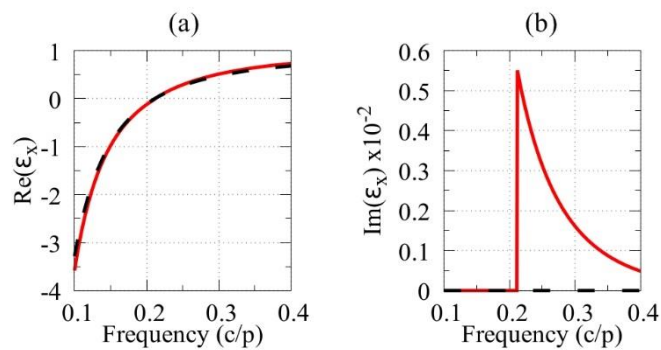

Fig.3. Real part (a), and imaginary part (b) of effective $\varepsilon_x$ of a FM with parameters $h = 0.001p$, $f_h = 0.1$, $b=a=0.9p$, and $\varepsilon_h = \varepsilon_d = 1$ from EMA (black dashed line) and QFS (red solid line).

Next, we discuss the effective $\varepsilon_x$ of a FM. The hole-array layer height and its filling ratio are defined as $h = 0.001p$ and $f_h = 0.1$, respectively. Figure 3(a)-(b) show real part and imaginary part of effective $\varepsilon_x$, respectively, as a function of frequency. The $\varepsilon_x$ from both EMA and QFS methods are real values and perfectly consistent at frequencies below the spoof plasma frequency $f_p = 0.211\, c/p$. The real value of $\varepsilon_x$ arises from the real $k_z$ at given frequency and $k_x = \pi/p$. The origin of the real $k_z$ will be shortly explained. However, the imaginary part of $\varepsilon_x$ from QFS method appears at a frequency above the spoof plasma frequency while that from EMA remains zero as shown in Fig. 3(b). The imaginary part of effective $\varepsilon_x$ from QFS exponentially decreases by increasing frequency. The complex value of $\varepsilon_x$ originates from the complex $k_z$ at given frequency and $k_x = 0$ which will become apparent when we discuss its EFC. Therefore, our effective parameters are applicable only for frequencies below the spoof plasma frequency. However, they cannot explain losses which occur at frequencies above the spoof plasma frequency.

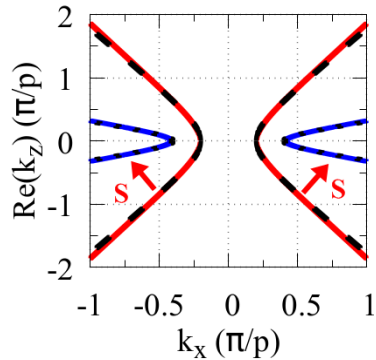

Fig. 4. EFCs of a FM at frequency 0.1 $c/p$ from QFS (red solid line) and EMA (black dashed line), and at frequency 0.2 $c/p$ from QFS (blue solid line) and EMA (black dotted line). The arrows indicate Poynting vector **S**. The parameters of FM are the same as those in Fig. 3.

Next, we discuss the EFCs of a FM. Figure 4 shows the EFCs of a FM whose parameters are the same as those in Fig. 3 at frequencies 0.1 $c/p$ and 0.2 $c/p$. These two frequencies lie below the spoof plasma frequency and thus exhibit the type-II hyperbolic EFCs. It can be seen that both models give consistent EFCs with zero imaginary part of $k_z$. The imaginary part of $k_z$ is zero because admittances in all regions are complex numbers at $k_x > k_0$ which are matched to the complex admittance of a hole-array layer giving purely real $r$ and $t$ coefficients. We employ $k_z$ at $k_x = \pi/p > k_0$ to compute effective $\varepsilon_x$ of an hyperbolic FM, and thus the effective $\varepsilon_x$ is purely real at a frequency below the spoof plasma frequency. These effective media also show the negative refraction in which the directions of Poynting vector and wavevector along $z$ axis are opposite [7]. At lower frequencies, $k_z$ are higher and the Poynting vector tend to be parallel to the $x$ axis. This means that the energy does not transmit through the structure in the extremely deep-subwavelength scale.

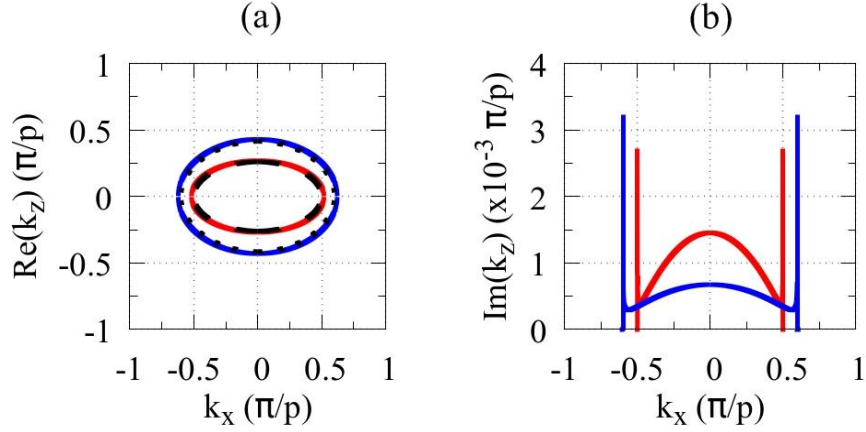

Fig. 5. Real part (a), and imaginary part (b) of EFCs of a FM at frequency 0.25 $c/p$ from QFS (red solid line) and EMA (black dashed line), and at frequency 0.3 $c/p$ from QFS (blue solid line) and EMA (black dotted line). The parameters of the FM are the same as those in Fig. 3.

Above the spoof plasma frequency, this same structure becomes elliptical medium. Figure 5(a) shows the elliptical EFCs at frequencies 0.25 $c/p$ and 0.3 $c/p$ using only real part of $k_z$ from QFS and EMA. It can be seen that the elliptical EFCs from EMA are perfectly consistent with the elliptical EFCs from QFS with only real part of $k_z$. The ellipse is larger by increasing frequency indicating the positive refraction in this medium. Figure 5(b) shows the imaginary part of $k_z$ as a function of $k_x$ from QFS (EMA has no imaginary part of $k_z$.) The imaginary contours are also in elliptical shape with the divergences at $k_x = k_0$ ( $k_0 = 0.5$ $\pi/p$ at $f = 0.25$ $c/p$ and $k_0 = 0.6$ $\pi/p$ at $f = 0.3$ $c/p$ ) due to the Wood's anomaly [19]. The imaginary part of $k_z$ at $k_x < k_0$ is lower by increasing the frequency. The imaginary part of $k_z$ arises because admittances in all regions except a hole-array layer are real which are mismatched to the complex admittance of a hole-array layer giving complex $r$ and $t$ coefficients. The disagreement on EFCs between both methods prohibit the application of the EMA in this frequency regime.

## 2. Subwavelength focusing

Finally, we theoretically demonstrate the application of a FM homogenized by EMA to behave as type-II HMM. The direct application of a type-II HMM is deep-subwavelength focusing where the energy focusing relies on the directional propagations of waves inside the HMM [6]. The large momentum mismatch between wave propagations in the HMM and wave propagations in its environment forces us to place a metallic grating or infinite slit array above the HMM. The slit array diffracts the incoming light into Bloch waves each denoted by

$\langle \mathbf{r} | \beta_m^{(x)} \rangle = \exp(i\beta_m^{(x)} x)/\sqrt{p_s}$, where $p_s$ is period of the slit array and $\beta_m^{(x)} = k_x + m2\pi/p_s$ where $m$ is an integer. The reflected light, the transmitted light, and the internal light inside the HMM are now expanded in terms of these Bloch waves. The $m$th-order admittance is now defined as $Y_m = \varepsilon k_0 / q_m$ where $\varepsilon = \varepsilon_{in}$ for input medium, $\varepsilon = \varepsilon_{out}$ for output medium, $\varepsilon = \varepsilon_x$ for HMM, and $q_m$ is $z$-component of wavevector determined by the dispersion relation of each medium. Unlike a hole, a slit has no cutoff frequency and its most dominant waveguide mode is TEM mode [20]. Therefore, the normalized slit waveguide mode denoted as $\langle \mathbf{r} | s \rangle$ is a constant and equal to $\langle \mathbf{r} | s \rangle = 1/\sqrt{w_s}$, where $w_s$ is slit width. Then, the coupled-mode analysis can be applied to this slit array/FM composite structure to obtain the field coefficients in all regions [21].

To show the deep-subwavelength focusing, we use the same parameters of a FM acting as type-II HMM as those used in Fig. 3 and 4, but now we work at the wavelength $25p$. For example, by choosing $p = 100\ \mu$m, then $h = 0.1\ \mu$m, $b = a = 90\ \mu$m, and $\lambda = 2,500\ \mu$m or frequency 0.12 THz. A metallic hole-array layer may be prepared by the atomic layer deposition followed by electron beam lithography, or by simpler ink-jet printing method. The $p_z$ obtained from $h$ and $f_h$ is equal to $0.01p$ satisfying the EMA's requirement. The hyperbolic EFC of the HMM at $\lambda = 25p$ calculated by the EMA as indicated by the red solid line in Fig. 6(a) clearly shows that wavevectors supported by the HMM are much larger than those supported by air as indicated by the black solid line. The angles of energy propagations with respect to $z$ axis inside the HMM, which is obtained by $\theta = \tan^{-1}(\varepsilon_x k_x / \varepsilon_z k_z)$, are shown in Fig. 6(b). The energy-propagation angles of large-wavevector waves lie in the narrow band 79.5°-82.2° which are usually referred to the critical angles of energy propagations denoted by $\theta_c$ [6]. Waves with $k_x$ close to $\sqrt{\varepsilon_d} k_0$ do not contribute to the focusing because $\theta$ jumps to 90° and thus the corresponding energy propagates parallel to the $x$ axis.

The air-filled PEC slit array with period $p_s = 10.0p$, slit width $w_s = 0.5p$, and slit height $h_s = 0.1p$ is then placed on the HMM in order to excite the large-wavevector waves. If we want to focus the energy at one spot between two slits, the number of supperlattices denoted by $N_{latt}$ should be chosen by $N_{latt} = \text{int}(p_s / 2p_z \tan \theta_c)$. By using $\theta_c = 79.5°$, we obtain $N_{latt} = 92$ ($N_{latt} p_z = 0.92p$) which is used as the starting point to optimize the correct number of supperlattices giving the best focusing spot as explained in the supplementary material. Figure 6(c) shows the normalized time-averaged Poynting vector behind the slit array of this structure with the optimized number of supperlattices $N_{latt} = 101$ ($N_{latt} p_z = 1.01p$). Two arms of directional energy propagations are clearly seen inside the HMM as expected. The energy is focused at the distance 0.959$p$ from the slit array which is close to the distance predicted by the EMA. The intensity at this spot is 285 times larger than the incident intensity which is also about 71 times larger than the near-field intensity in air behind the bare slit array without the HMM as shown in Fig. 6(d). The full width at half maximum along $x$ axis (FWHM$_x$) inside the HMM is 0.3$p$=$\lambda$/83 which is about 2.5 smaller than that in air without the HMM. However, only small portion of the focused energy is transmitted to the output medium, which is defined as air, due to the large impedance mismatch. At $z=L$ inside the output medium, the peak of transmitted intensity is only 30 times higher than the incident intensity with the FWHM$_x$ equal to 0.75$p$=$\lambda$/33 comparable to that of the bare rod array without HMM. The energy also decays in air as expected with the decay length 0.275$p$ because there is no magnification process [5] to reduce wavevectors of large-wavevector waves inside the flat HMM so that the transmitted waves have large wavevectors not supported in air. This effect is more severe with larger-wavevector waves which more quickly decay in air.

Fig. 6. (a) EFC of a HMM (red line) and air (black line) at $\lambda = 25p$. (b) Critical angles of energy propagations

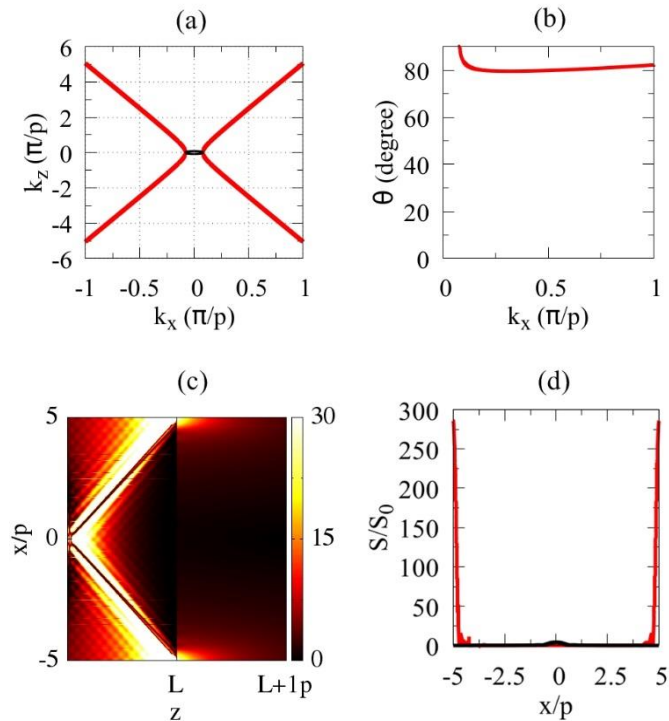

inside a HMM as a function of $k_x$ .(c) Distribution of normalized time-averaged Poynting vector behind the slit array for a HMM with the number of supperlattices 101, where L denotes total length of the slit array/HMM structure. (d) Profile of normalized time-averaged Poynting vector at distance 0.959$p$ from the slit array. The red line is the profile inside the slit array/HMM structure, and the black line is the profile in air without the HMM. All figures are obtained by EMA method.

Subwavelength focusing is particularly important in optical nano-imaging. A. J. Huber et al have experimentally utilized metalized-AFM tip which localizes the incoming terahertz light into nano-volume around the tip's apex to probe nanoscale features with the spatial resolution down to 40 nm at 2.54 THz ($\lambda/2,950$) [22]. This spatial resolution is far better than ours. However, the slit-array/HMM structure has flat surface which is more practical than the curve surface of the tip's apex. The solutions of electromagnetic waves in our structure is also analytically known unlike those in the tip which are computed by numerical method. If the spatial resolution, the $FWHM_x$, of the slit-array/HMM can be further reduced and the out-coupling mechanism [23] is employed to extract the focused energy inside the HMM, this structure may become alternative choice of the metallized-AFM tip for subwavelength imaging. $FWHM_x$ reduction may be done, for example, by lowering the spoof plasma frequency or decreasing the slit width. Our effective parameters may be used to aid these designs and to build more complicated FM-based composite structure.

**Conclusion**

In summary, the closed-form formulae of effective parameters of stacked hole-array layers known as the fishnet metamaterial have been formulated for light wavelength much larger than hole-array period. The anisotropic fishnet metamaterial behaves like non-magnetic insulator medium along longitudinal axis, but it behaves like metal along transverse axis. The effective medium approximation of a hyperbolic fishnet metamaterial is perfectly consistent with the quasi-full solution method using the homogenized hole-array layer when the supperlattice period is equal to or smaller than 0.01 times the hole-array period. We also theoretically demonstrate the application of our effective parameters in subwavelength focusing using the composite structure of slit array and fishnet metamaterial. It is found that the focused intensity inside the fishnet metamaterial is about 71 times larger than the maximum intensity on the same plane in air without the fishnet metamaterial, and the full width at half maximum corresponding to the former is $\lambda/83$ which is about 2.5 times smaller than the latter. The energy is strongly reflected at the fishnet metamaterial's end interface due to the large difference of wavevectors reducing the intensity and widening the full width at half maximum of the transmitted light.

**Supplementary material**

See supplementary material for full details of QFS method and optimization process of the subwavelength focusing.

**Acknowledgement**

This work is supported by King Monkut's Institute of Technology Ladkrabang [KREF145905].

## Supplementary material

### 1. Quasi-full solution (QFS) method

The transmission coefficients ($t$) and the reflection coefficient ($r$) are obtained by imposing the continuities of $E_x$ and $H_y$ at all interfaces to apply the transfer matrix method. Then, the coefficients $t$ and $r$ are written as follows

$$t = -S_{21}/S_{22}, \quad r = (S_{11}S_{22} - S_{12}S_{21})/S_{22}, \tag{S1}$$

where $S_{ij}$ is a matrix element of the scattering matrix defined by $\mathbf{S} = \mathbf{T}_{\text{out}} \mathbf{T}_{\text{d,h}} \mathbf{T}_{\text{in}}$, where the transfer matrices $\mathbf{T}_{\text{in}}$, $\mathbf{T}_{\text{d,h}}$, and $\mathbf{T}_{\text{out}}$ are given as follows

$$\mathbf{T}_{\text{in}} = \frac{1}{\tau_{\text{h,in}}} \begin{bmatrix} 1 & \rho_{\text{h,in}} \\ \rho_{\text{h,in}} & 1 \end{bmatrix}, \quad \mathbf{T}_{\text{d,h}} = \frac{1}{\tau_{\text{d,h}}} \begin{bmatrix} e^{i\phi_{01}^{(h)}} & \rho_{\text{d,h}} e^{-i\phi_{01}^{(h)}} \\ \rho_{\text{d,h}} e^{i\phi_{01}^{(h)}} & e^{-i\phi_{01}^{(h)}} \end{bmatrix}, \quad \mathbf{T}_{\text{out}} = \begin{bmatrix} \tau_{\text{d,out}} e^{i\phi_d} & 0 \\ -\rho_{\text{d,out}} e^{i\phi_d} & e^{-i\phi_d} \end{bmatrix}. \tag{S2}$$

### 2. Optimization process of subwavelength focusing

The best subwavelength focusing should give the highest focused intensity and good intensity contrast like the one shown in Fig. 6(c) and (d) of the main text. This critically depends on thickness of a hyperbolic metamaterial (HMM). We optimize the composite structure by fixing the slit width $w_s = 0.5p$ and the slit height $h_s = 0.1p$, but the slit period $p_s$ is varied from $p_s = 0.8p$ to $p_s = 18.0p_s$. This means that the initial HMM thickness determined by $N_{\text{latt}} p_z$ is changed following the initial $N_{\text{latt}}$ which is determined by the $p_s$ using the relation $N_{\text{latt}} = \text{int}(p_s / 2p_z \tan\theta_c)$ with $\theta_c = 79.5°$ and $p_z = 0.01p$. For each $p_s$, we then increase the value of $N_{\text{latt}}$ starting from the initial $N_{\text{latt}}$ and compute the corresponding normalized intensity profiles along the $x$ axis at $z = L = h_s + N_{\text{latt}} p_z$ in the output medium until the best $N_{\text{latt}}$ is found. In the intensity calculation, all diffraction modes starting from $m = 0, \pm 1, \pm 2, \ldots$ are included until the intensity is converged. The focused intensity enhancement $S/S_0$ is then computed at the points $x = \pm p_s/2$ in the optimized intensity profile. Figure S1(a) and (b) shows the optimized $N_{\text{latt}}$ and the corresponding focused intensity enhancement $S/S_0$, respectively, as a function of $p_s$. Figure S1(a) shows that the optimized $N_{\text{latt}}$ linearly increases by increasing $p_s$. The peak of the focused intensity enhancement $S/S_0$ in Fig. S1(b) occurs at $p_s = 10.0p$. By using the optimized $N_{\text{latt}}$ for each $p_s$ from Fig. S1(a), we compute the focused intensity enhancement $S/S_0$ inside the HMM by finding the maximum of the intensity enhancement in the intensity profiles along the lines $x = \pm p_s/2$. The focused intensity enhancement inside the HMM shown in Fig. S1(c) clearly shows the peak position at $p_s = 10.0p$ consistent with the peak position in Fig. S1(b). Therefore, we can rapidly optimize the best subwavelength focusing by optimizing the intensity enhancement profiles along the $x$ axis at $z=L$ in the output medium.

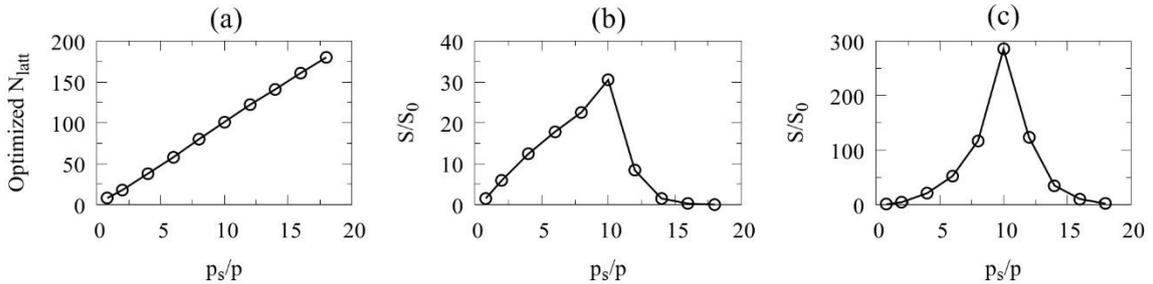

Fig. S1. Optimized $N_{\text{latt}}$ (a), focused intensity enhancement in the output medium at $z=L$ (b), and focused intensity enhancement in the HMM (c) as a function of $p_s$.